\documentclass[12pt,preprint]{aastex}

\def\msun{\mbox{M}_{\odot}}

\shorttitle{Correlated Variability in Mkn 509}
\shortauthors{Marshall \& Miller}

\begin{document}

\title{Correlated X-ray and Optical Variability in Mkn~509}
\author{Kevin Marshall\altaffilmark{1}\altaffilmark{2}, Wesley T.~Ryle\altaffilmark{2}, H.~Richard Miller\altaffilmark{2}}
\altaffiltext{1}{Department of Physics, Bucknell University, Lewisburg PA 17837, {\tt kevin.marshall@bucknell.edu}}
\altaffiltext{2}{Department of Physics and Astronomy, Georgia State University, Atlanta, GA 30303}

\begin{abstract}
We present results of a 3 year monitoring campaign of the Seyfert~1 galaxy
Markarian~509, using X-ray data from the {\em Rossi X-ray Timing Explorer} and
optical data taken by the SMARTS consortium.  Both light curves show significant
variations, and are strongly correlated with the optical flux leading the X-ray
flux by 15 days.  The X-ray power spectrum shows a steep high-frequency slope
of $-2.0$, breaking to a slope of $-1.0$ at at timescale of 34 days.  The lag
from optical to X-ray emission is most likely caused by variations in the accretion
disk propagating inward.
\end{abstract}


\section{Introduction}

Variability has long been recognized as one of the defining characteristics of
active galactic nuclei (AGN).  This variability is not limited to one wavelength
regime, but is spread across the entire electromagnetic spectrum.  Produced in
the innermost regions of the nucleus are X-ray and optical/UV continuum photons.
Current models involve an accretion disk emitting thermal ($10^5$~K) photons in
the optical and UV range \citep{Malkan1983,Shields1978}, which are then inverse Compton scattered to X-ray
energies by a corona of hot electrons above the accretion disk \citep{Haardt1991,Haardt1993}.

By monitoring the variability at optical and X-ray wavelengths, we can learn
information about the geometry and exact mechanism responsible for the emission.
The traditional view is that because the X-rays are produced closer to the
nucleus, variations should be the largest at that wavelength, with optical
variations lagging behind.  On the other hand, the variations may originate in
the optical seed photons, which then cause the X-ray emission to lag behind the
optical.

Previous attempts at correlating X-ray and optical variability in Seyfert galaxies have been
inconclusive.  On long timescales, \citet{Uttley2003} detected a strong
correlation in NGC~5548 with no lag, while \citet{Chiang2000} found a weaker
correlation on short timescales.  The same was true for NGC~4051, with
\citet{Peterson2000} finding a correlation on long but not short timescales.
However this is not the case with all Seyferts, as \citet{Maoz2002} found no
correlation in NGC~3516 after 5 years of monitoring.  \citet{Shemmer2003} and
\citet{Arevalo2005} both found correlations on short timescales in NGC~4051 and
MCG--6-30-15, respectively, while \citet{Papadakis2000} found no correlation
on short timescales in Akn~564.  For a more comprehensive review of past
efforts, see \citet{Maoz2002} and references therein.

Mkn~509 is a nearby ($z=0.034$) Seyfert~1 galaxy.  First detected in the
X-rays by {\em Ariel V} \citep{Cooke1978}, Mkn~509 was later found to have a
soft X-ray excess by \citet{Singh1989}.  Simultaneous observations with
{\em GINGA} and {\em ROSAT} again showed a soft excess, with some flattening at
higher energies due to reflection \citep{Pounds1994}.  Later observations with
{\em ASCA} showed the presence of an Fe~K$\alpha$ line at 6.4~keV, and the
presence of a warm absorber rather than a soft excess \citep{Reynolds1997}.
More recent observations with {\em XMM-Newton} show the origin of the soft X-ray
excess lies mostly in thermal emission from the inner parts of the accretion
disk \citep{Pounds2001}.

We present here results from a 3.5 year X-ray and optical monitoring campaign
on Mkn~509.  In \S2, we show light curves and discuss data analysis methods
for all of our observations.  In \S3, we calculate the X-ray power density
spectrum (PDS), using Monte Carlo methods.  In \S4, we present results of
cross-correlation between the X-ray and optical light curves and discuss the
significance of the results.  Finally, in \S5 we compare our result to previous
efforts, and discuss possible implications for physical models of accretion.

\section{Observations and Data Reduction}

\subsection{X-ray Data}

Mkn~509 was observed with the {\em Rossi X-ray Timing Explorer (RXTE)} from
28 March 2003 -- 29 June 2006 in X-rays, with sampling every 3 days.  We
use only data taken by PCUs 0 and 2, in STANDARD2 data mode.  All
of our data were reduced using FTOOLS~v5.2 software, provided by HEASARC.
Data were excluded if the Earth elevation angle was $< 15^{\circ}$, pointing
offset $> 0.02^{\circ}$, time since South Atlantic Anomaly passage $< 30$
minutes, or electron noise $> 0.01$ units.  Counts were extracted from the top
PCU layer only to maximize the signal to noise ratio.  

\subsection{Optical Data}

Ground-based optical observations were taken roughly twice per week  from 23~October~2003 --
12~November~2006, as weather conditions and proximity to the sun allowed. 
All data were taken with the 1.3m telescope at Cerro Tololo
Inter-American Observatory (CTIO), operated by the SMARTS consortium.  The 1.3m
reflector uses the ANDICAM instrument, which has a Fairchild 447
$2048 \times 2048$ CCD chip with 15~$\mu$m pixels.  Using $2 \times 2$ binning,
this yields a 6-arcminute field of view, with a plate scale of
$0.369^{\prime\prime}$ per pixel.  Observations were taken in standard Johnson
$BVR$ filters; however we present only the $R$ filter data here.

Image processing was done with IRAF, using standard methods.  A large number
(typically 10) of bias and flat fields were taken each night, and then combined
using a min/max rejection algorithm in IRAF.

Photometry was done using the CCDPHOT program, developed for IDL by Marc Buie.
A $7^{\prime\prime}$ aperture was chosen to minimize any contributions from the
host galaxy.  Photometric measurements were taken for the object and check stars
A and F from \citet{Miller1981}.  We use differential photometry, i.e. no
standard stars were used.  Object brightness was calculated by comparison with
check star A, and error bars are given by the standard deviation of the
difference in brightness between check stars A and F.

\section{Light Curves and the Power Density Spectrum}

The X-ray and optical light curves are shown in Figure~1, and optical magnitudes
are given in Table~1.  The X-ray data are
more complete, owing to weather and other observing constraints with
ground-based telescopes.  The X-ray data also show more high-frequency
variability than the optical data.

To begin to quantify the intrinsic variability of each light curve, we have
calculated the fractional variability, defined as

\[ F_{\rm var} = \sqrt{\frac{S^2 - \langle \sigma_{\rm err}^2 \rangle}{\langle \mu \rangle^2}} \]

where $S^2$ is the variance of the light curve, $\langle \sigma_{\rm err}^2 \rangle$ is the mean error squared, and $\langle \mu \rangle$ is the mean count
rate \citep{Markowitz2004}.  For Mkn~509, we find a fractional variability of
$F_{\rm var} = 19$\% and 12\% for the X-ray and optical light curves,
respectively.  Note that we do not subtract any host galaxy flux from the
optical data, so this figure represents a lower limit on the true value of
$F_{\rm var}$.

We use the Monte Carlo method of \citet{Uttley2002} and \citet{Markowitz2003} to calculate
the power density spectrum (PDS).  Briefly, a long light curve is simulated
using the method of \citet{Timmer1995}.  This light curve is then split into
parts, and the PDS calculated for each segment.  The individual power spectra
are then averaged, and then compared to the observed power spectrum.  This is
done over a grid of slopes and break frequencies.

For Mkn~509, we fixed the low frequency slope of the PDS at $-1.0$, breaking
to a steeper slope at some break frequency, $\nu_{\rm b}$.  We use
high-frequency slopes between $-1.0$ and $-2.5$ for our input models, and
break frequencies between $-8.0 \le \nu_{\rm b} \le -5.5$~Hz, incrementing the slope
by 0.1 and the break frequency by factors of 1.5 (0.18 in the logarithm).

The simulation results are shown in Figure~2.  The best fit occurs at a high
frequency slope of $-2.0$ and a break frequency of $\nu_{\rm b}=-6.47$~Hz, or
a timescale of 34~days.  The probability of acceptance of this 95\%.  For an
unbroken power law, the probability of acceptance was only 29\%, with a single
slope of $-1.2$.

Note that the PDS covers only slightly more than 2 decades in frequency,
due to the lack of additional high-frequency data.  We have proposed
for more intensive observations with {\em RXTE}, which will allow us to fill in the
high-frequency area of the PDS.  Until then, the break timescale of 34~days
should be regarded more as an upper limit and not a concrete value.

A break timescale less than 34~days does agree well with the mass-break
frequency relationship discussed by \citet{Uttley2005} and \citet{Markowitz2003}.  With a black hole mass of $1.43 \times 10^8 \msun$ \citep{Peterson2004}, we
would expect to find a break timescale of $\tau \sim 10-50$~days.  Again, more
high-frequency observations are needed to better constrain this value.

\section{Cross Correlation Function}

To examine the possibility of a correlation between X-ray and optical flux, we
use the cross correlation function (CCF).  The traditional CCF requires evenly
sampled data, and can be computationally intensive.  Because of observing
constraints, neither our X-ray nor optical data are evenly sampled.  To solve
this issue, we use the discrete correlation function (DCF) of \citet{Edelson1988}, which allows for cross-correlation of two unevenly sampled data sets.

The DCF is shown in Figure~3, with the convention that positive lag indicating optical variations
leading the X-rays.  The maximum value occurs at a lag of $\tau = +15$~days,
with a correlation coefficient of $r=0.93$.

As discussed in \citet{Uttley2003}, traditional error bars are inadequate for
assessing the significance of the DCF, because adjacent data points in the
light curve are ``red noise'' data and not uncorrelated.  Therefore, similar to
\citet{Uttley2003}, we use Monte Carlo simulations to test the significance of
both the correlation and measured lag in our DCF.

We began by simulating 2 independent red noise light curves, using the method of
\citet{Timmer1995}.  Both light curves were given underlying model power
spectra with slope $-1$ at low frequencies, breaking to a slope of $-2$ above
a break frequency of $\log \nu = -6.47$.  The two light curves were then
re-sampled in the same fashion as the original data, and random noise was added
in the form of a Gaussian random with mean of zero and standard deviation equal
to the average observed error.  For $10^4$ simulations, we found a maximum
correlation coefficient $r > 0.93$ only 38 times.  Therefore the correlation
coefficient of $r=0.93$ seen in the data is significant at more than 99\%
confidence.

To test the significance of the measured lag, we use simulations similar to
above.  In this case, 2 identical light curves were generated, with one
lagging 15 days behind the other.  The simulated data were then re-sampled in
the same fashion as the observed light curves, and random noise was added
using the same method as above.  We then searched for cases where the peak
lag was greater than the lag at 15 days.  Out of $10^4$ simulations, we found
only 287 cases where the peak lag did not occur at $\tau = 15$~days.  Therefore
our measured lag is significant at more than 97\% confidence.

\section{Conclusions}

We have shown that over long timescales, optical variations lead the X-rays by
15~days for Mkn~509.  Initially, this lag would appear to be far too large for reprocessing
models where thermal optical/UV photons from the disk are inverse Compton
scattered to X-ray energies \citep{Haardt1991}.

However, such a lag is not unprecedented.  \citet{Shemmer2003} found an
optical to X-ray lag of 2.1~days in NGC~4051.  Such a lag is most likely related
to the viscous or thermal timescale.  In that case, as variations in the
accretion flow propagate inwards through the disk, they pass first through the
optical emitting region, and then later through the X-ray emitting region
closer to the black hole.  The additional high-frequency variability seen in
the X-rays could come from the decreased light-travel time closer to the black
hole, or from an additional emission component closer to the central engine.

If the lag is due to changes in the accretion flow, wouldn't we expect to find
such a lag in all Seyfert galaxies?  Probably not, since the mass of the black
hole in Mkn~509 is $1.43 \times 10^{8} \msun$ \citep{Peterson2004}, roughly an
order of magnitude greater than previously studied objects.  Many of the
relevant timescales (orbital, thermal, viscous) increase linearly with mass, so
we would expect to see lags of 1-2 days for other, less massive Seyferts.

Previous observing campaigns have involved weekly monitoring for a period of
years, or more intensive monitoring for a period of a few days.  In those cases,
it is distinctly possible that the data would appear to be correlated, but with
zero lag if the monitoring is on a weekly basis.  Conversely, the data would
appear to be uncorrelated if the lag time is greater than the monitoring period
of just a few days.  By observing a more massive galaxy on a frequent basis for
several years, we have been able to measure a lag between the optical and X-ray
emission.

\acknowledgements{We thank the referee for a useful report, which improved the readability of this paper.  KM, WRT, and HRM acknowledge support from NASA grant
No. NNG-04-G04-6G, and from the PEGA program at GSU.}

\bibliographystyle{apj}

\bibliography{ms}

\begin{thebibliography}{24}
\expandafter\ifx\csname natexlab\endcsname\relax\def\natexlab#1{#1}\fi

\bibitem[{{Ar{\'e}valo} {et~al.}(2005){Ar{\'e}valo}, {Papadakis}, {Kuhlbrodt},
  \& {Brinkmann}}]{Arevalo2005}
{Ar{\'e}valo}, P., {Papadakis}, I., {Kuhlbrodt}, B., \& {Brinkmann}, W. 2005,
  \aap, 430, 435

\bibitem[{{Chiang} {et~al.}(2000){Chiang}, {Reynolds}, {Blaes}, {Nowak},
  {Murray}, {Madejski}, {Marshall}, \& {Magdziarz}}]{Chiang2000}
{Chiang}, J., {Reynolds}, C.~S., {Blaes}, O.~M., {Nowak}, M.~A., {Murray}, N.,
  {Madejski}, G., {Marshall}, H.~L., \& {Magdziarz}, P. 2000, \apj, 528, 292

\bibitem[{{Cooke} {et~al.}(1978){Cooke}, {Ricketts}, {Maccacaro}, {Pye},
  {Elvis}, {Watson}, {Griffiths}, {Pounds}, {McHardy}, {Maccagni}, {Seward},
  {Page}, \& {Turner}}]{Cooke1978}
{Cooke}, B.~A., {Ricketts}, M.~J., {Maccacaro}, T., {Pye}, J.~P., {Elvis}, M.,
  {Watson}, M.~G., {Griffiths}, R.~E., {Pounds}, K.~A., {McHardy}, I.,
  {Maccagni}, D., {Seward}, F.~D., {Page}, C.~G., \& {Turner}, M.~J.~L. 1978,
  \mnras, 182, 489

\bibitem[{{Edelson} \& {Krolik}(1988)}]{Edelson1988}
{Edelson}, R.~A., \& {Krolik}, J.~H. 1988, \apj, 333, 646

\bibitem[{{Haardt} \& {Maraschi}(1991)}]{Haardt1991}
{Haardt}, F., \& {Maraschi}, L. 1991, \apjl, 380, L51

\bibitem[{{Haardt} \& {Maraschi}(1993)}]{Haardt1993}
---. 1993, \apj, 413, 507

\bibitem[{{Malkan}(1983)}]{Malkan1983}
{Malkan}, M.~A. 1983, \apj, 268, 582

\bibitem[{{Maoz} {et~al.}(2002){Maoz}, {Markowitz}, {Edelson}, \&
  {Nandra}}]{Maoz2002}
{Maoz}, D., {Markowitz}, A., {Edelson}, R., \& {Nandra}, K. 2002, \aj, 124,
  1988

\bibitem[{{Markowitz} \& {Edelson}(2004)}]{Markowitz2004}
{Markowitz}, A., \& {Edelson}, R. 2004, \apj, 617, 939

\bibitem[{{Markowitz} {et~al.}(2003){Markowitz}, {Edelson}, {Vaughan},
  {Uttley}, {George}, {Griffiths}, {Kaspi}, {Lawrence}, {McHardy}, {Nandra},
  {Pounds}, {Reeves}, {Schurch}, \& {Warwick}}]{Markowitz2003}
{Markowitz}, A., {Edelson}, R., {Vaughan}, S., {Uttley}, P., {George}, I.~M.,
  {Griffiths}, R.~E., {Kaspi}, S., {Lawrence}, A., {McHardy}, I., {Nandra}, K.,
  {Pounds}, K., {Reeves}, J., {Schurch}, N., \& {Warwick}, R. 2003, \apj, 593,
  96

\bibitem[{{Miller}(1981)}]{Miller1981}
{Miller}, H.~R. 1981, \aj, 86, 87

\bibitem[{{Papadakis} {et~al.}(2000){Papadakis}, {Brinkmann}, {Negoro},
  {Detsis}, {Papamastorakis}, \& {Gliozzi}}]{Papadakis2000}
{Papadakis}, I.~E., {Brinkmann}, W., {Negoro}, H., {Detsis}, E.,
  {Papamastorakis}, I., \& {Gliozzi}, M. 2000, ArXiv Astrophysics e-prints

\bibitem[{{Peterson} {et~al.}(2004){Peterson}, {Ferrarese}, {Gilbert}, {Kaspi},
  {Malkan}, {Maoz}, {Merritt}, {Netzer}, {Onken}, {Pogge}, {Vestergaard}, \&
  {Wandel}}]{Peterson2004}
{Peterson}, B.~M., {Ferrarese}, L., {Gilbert}, K.~M., {Kaspi}, S., {Malkan},
  M.~A., {Maoz}, D., {Merritt}, D., {Netzer}, H., {Onken}, C.~A., {Pogge},
  R.~W., {Vestergaard}, M., \& {Wandel}, A. 2004, \apj, 613, 682

\bibitem[{{Peterson} {et~al.}(2000){Peterson}, {McHardy}, {Wilkes}, {Berlind},
  {Bertram}, {Calkins}, {Collier}, {Huchra}, {Mathur}, {Papadakis}, {Peters},
  {Pogge}, {Romano}, {Tokarz}, {Uttley}, {Vestergaard}, \&
  {Wagner}}]{Peterson2000}
{Peterson}, B.~M., {McHardy}, I.~M., {Wilkes}, B.~J., {Berlind}, P., {Bertram},
  R., {Calkins}, M., {Collier}, S.~J., {Huchra}, J.~P., {Mathur}, S.,
  {Papadakis}, I., {Peters}, J., {Pogge}, R.~W., {Romano}, P., {Tokarz}, S.,
  {Uttley}, P., {Vestergaard}, M., \& {Wagner}, R.~M. 2000, \apj, 542, 161

\bibitem[{{Pounds} {et~al.}(2001){Pounds}, {Reeves}, {O'Brien}, {Page},
  {Turner}, \& {Nayakshin}}]{Pounds2001}
{Pounds}, K., {Reeves}, J., {O'Brien}, P., {Page}, K., {Turner}, M., \&
  {Nayakshin}, S. 2001, \apj, 559, 181

\bibitem[{{Pounds} {et~al.}(1994){Pounds}, {Nandra}, {Fink}, \&
  {Makino}}]{Pounds1994}
{Pounds}, K.~A., {Nandra}, K., {Fink}, H.~H., \& {Makino}, F. 1994, \mnras,
  267, 193

\bibitem[{{Reynolds}(1997)}]{Reynolds1997}
{Reynolds}, C.~S. 1997, \mnras, 286, 513

\bibitem[{{Shemmer} {et~al.}(2003){Shemmer}, {Uttley}, {Netzer}, \&
  {McHardy}}]{Shemmer2003}
{Shemmer}, O., {Uttley}, P., {Netzer}, H., \& {McHardy}, I.~M. 2003, \mnras,
  343, 1341

\bibitem[{{Shields}(1978)}]{Shields1978}
{Shields}, G.~A. 1978, \nat, 272, 706

\bibitem[{{Singh} {et~al.}(1989){Singh}, {Westergaard}, {Schnopper}, {Awaki},
  \& {Tawara}}]{Singh1989}
{Singh}, K.~P., {Westergaard}, N.~J., {Schnopper}, H.~W., {Awaki}, H., \&
  {Tawara}, Y. 1989, in ESA Special Publication, Vol. 296, ESA Special
  Publication, ed. J.~{Hunt} \& B.~{Battrick}, 1053--1058

\bibitem[{{Timmer} \& {Koenig}(1995)}]{Timmer1995}
{Timmer}, J., \& {Koenig}, M. 1995, \aap, 300, 707

\bibitem[{{Uttley} {et~al.}(2003){Uttley}, {Edelson}, {McHardy}, {Peterson}, \&
  {Markowitz}}]{Uttley2003}
{Uttley}, P., {Edelson}, R., {McHardy}, I.~M., {Peterson}, B.~M., \&
  {Markowitz}, A. 2003, \apjl, 584, L53

\bibitem[{{Uttley} \& {McHardy}(2005)}]{Uttley2005}
{Uttley}, P., \& {McHardy}, I.~M. 2005, \mnras, 363, 586

\bibitem[{{Uttley} {et~al.}(2002){Uttley}, {McHardy}, \&
  {Papadakis}}]{Uttley2002}
{Uttley}, P., {McHardy}, I.~M., \& {Papadakis}, I.~E. 2002, \mnras, 332, 231

\end{thebibliography}

\begin{deluxetable}{cccc}
\tablewidth{0pt}
\tablecaption{Optical Data}
\tabletypesize{\scriptsize}
\tablehead{
\colhead{MJD\tablenotemark{a}} & \colhead{obj--chkA\tablenotemark{b}} & \colhead{obj--chkF\tablenotemark{b}} & \colhead{chkA--chkF\tablenotemark{b}}
}
                                                                                \startdata

52939.57464 & -1.152 & -3.165 & -2.013 \\
52942.55968 & -1.161 & -3.184 & -2.024 \\
52945.53301 & -1.185 & -3.197 & -2.012 \\
52948.54159 & -1.202 & -3.211 & -2.009 \\
52951.53280 & -1.212 & -3.233 & -2.021 \\
52954.52887 & -1.221 & -3.231 & -2.011 \\
52957.52418 & -1.241 & -3.243 & -2.002 \\
52961.51394 & -1.244 & -3.264 & -2.020 \\
52964.50961 & -1.240 & -3.322 & -2.082 \\
52968.51123 & -1.256 & -3.235 & -1.979 \\
52972.51042 & -1.257 & -3.158 & -1.901 \\
52975.51269 & -1.270 & -3.319 & -2.049 \\
52978.51179 & -1.226 & -3.478 & -2.253 \\
52982.50990 & -1.208 & -3.783 & -2.575 \\
53125.85882 & -1.142 & -3.157 & -2.015 \\
53130.87646 & -1.150 & -3.158 & -2.008 \\
53134.79642 & -1.148 & -3.149 & -2.001 \\
53142.88867 & -1.132 & -3.135 & -2.003 \\
53153.90404 & -1.092 & -3.101 & -2.009 \\
53157.88565 & -1.117 & -3.111 & -1.994 \\
53160.81833 & -1.113 & -3.117 & -2.004 \\
53174.87961 & -1.107 & -3.128 & -2.021 \\
53180.83470 & -1.111 & -3.120 & -2.009 \\
53188.82977 & -1.112 & -3.135 & -2.023 \\
53191.82883 & -1.104 & -3.129 & -2.025 \\
53194.81454 & -1.123 & -3.147 & -2.024 \\
53197.85733 & -1.150 & -3.160 & -2.010 \\
53206.83395 & -1.182 & -3.178 & -1.996 \\
53217.76241 & -1.194 & -3.139 & -1.945 \\
53223.79976 & -1.193 & -3.210 & -2.018 \\
53240.77539 & -1.193 & -3.199 & -2.006 \\
53243.72964 & -1.193 & -3.222 & -2.029 \\
53246.72076 & -1.151 & -3.300 & -2.149 \\
53248.71606 & -1.182 & -3.207 & -2.024 \\
53250.70876 & -1.182 & -3.192 & -2.010 \\
53254.71429 & -1.164 & -3.177 & -2.013 \\
53262.69859 & -1.153 & -3.160 & -2.006 \\
53265.71466 & -1.139 & -3.110 & -1.971 \\
53268.68162 & -1.149 & -3.171 & -2.022 \\
53269.73045 & -1.133 & -3.156 & -2.023 \\
53274.68698 & -1.143 & -3.208 & -2.064 \\
53278.63685 & -1.137 & -3.152 & -2.015 \\
53281.65461 & -1.158 & -3.169 & -2.012 \\
53289.59618 & -1.143 & -3.152 & -2.009 \\
53296.60557 & -1.139 & -3.151 & -2.012 \\
53298.59711 & -1.145 & -3.151 & -2.006 \\
53301.60686 & -1.170 & -3.176 & -2.006 \\
53307.60839 & -1.174 & -3.174 & -2.000 \\
53311.58837 & -1.173 & -3.185 & -2.012 \\
53324.53841 & -1.151 & -3.148 & -1.997 \\
53329.52019 & -1.153 & -3.176 & -2.023 \\
53560.85358 & -0.974 & -2.993 & -2.019 \\
53563.85279 & -0.949 & -2.959 & -2.010 \\
53570.74097 & -0.912 & -2.900 & -1.988 \\
53575.76890 & -0.868 & -2.879 & -2.011 \\
53578.82825 & -0.842 & -2.844 & -2.002 \\
53581.71973 & -0.831 & -2.858 & -2.028 \\
53584.72747 & -0.819 & -2.838 & -2.019 \\
53587.76697 & -0.828 & -2.849 & -2.021 \\
53588.73949 & -0.842 & -2.819 & -1.977 \\
53591.69279 & -0.847 & -2.836 & -1.990 \\
53599.76873 & -0.821 & -2.805 & -1.984 \\
53608.74477 & -0.814 & -2.812 & -1.998 \\
53618.72031 & -0.824 & -2.823 & -1.999 \\
53626.66931 & -0.832 & -2.851 & -2.019 \\
53633.60940 & -0.861 & -2.853 & -1.992 \\
53640.60154 & -0.890 & -2.826 & -1.936 \\
53644.59138 & -0.920 & -2.906 & -1.986 \\
53654.56947 & -0.916 & -2.924 & -2.008 \\
53661.58605 & -0.916 & -2.919 & -2.002 \\
53668.53228 & -0.945 & -2.945 & -2.000 \\
53676.52679 & -0.917 & -2.922 & -2.005 \\
53682.53441 & -0.888 & -2.923 & -2.035 \\
53704.52323 & -0.850 & -2.870 & -2.020 \\
53946.76870 & -0.988 & -2.979 & -1.991 \\
53951.66911 & -0.962 & -2.963 & -2.001 \\
53960.75079 & -0.982 & -2.988 & -2.006 \\
53963.71572 & -0.978 & -2.961 & -1.983 \\
53968.71505 & -0.947 & -2.951 & -2.004 \\
53970.75606 & -0.932 & -2.943 & -2.011 \\
53974.72391 & -0.922 & -2.933 & -2.011 \\
53979.69105 & -0.918 & -2.929 & -2.012 \\
53986.64399 & -0.947 & -2.854 & -1.907 \\
53989.68397 & -0.949 & -2.987 & -2.038 \\
53993.66439 & -0.982 & -2.986 & -2.004 \\
53998.61144 & -1.004 & -3.005 & -2.001 \\
54003.62051 & -1.035 & -3.028 & -1.993 \\
54007.64488 & -1.051 & -3.037 & -1.985 \\
54016.57288 & -1.085 & -3.086 & -2.002 \\
54019.57759 & -1.082 & -3.087 & -2.005 \\
54023.54125 & -1.069 & -3.081 & -2.012 \\
54026.54558 & -1.079 & -3.090 & -2.010 \\
54029.51316 & -1.077 & -3.079 & -2.001 \\
54032.54927 & -1.030 & -3.051 & -2.021 \\
54037.50093 & -1.061 & -3.082 & -2.022 \\
54040.51623 & -1.075 & -3.027 & -1.952 \\
54043.50410 & -1.075 & -3.077 & -2.002 \\
54046.51196 & -1.076 & -3.089 & -2.013 \\
54051.50853 & -1.075 & -3.096 & -2.021 \\

\enddata
\tablenotetext{a}{Modified Julian Date}
\tablenotetext{b}{Errors for photometry are given by the standard deviation of chkA -- chkF which is 0.026~mag.}

\end{deluxetable}

\begin{figure}
\figurenum{1}
\plotone{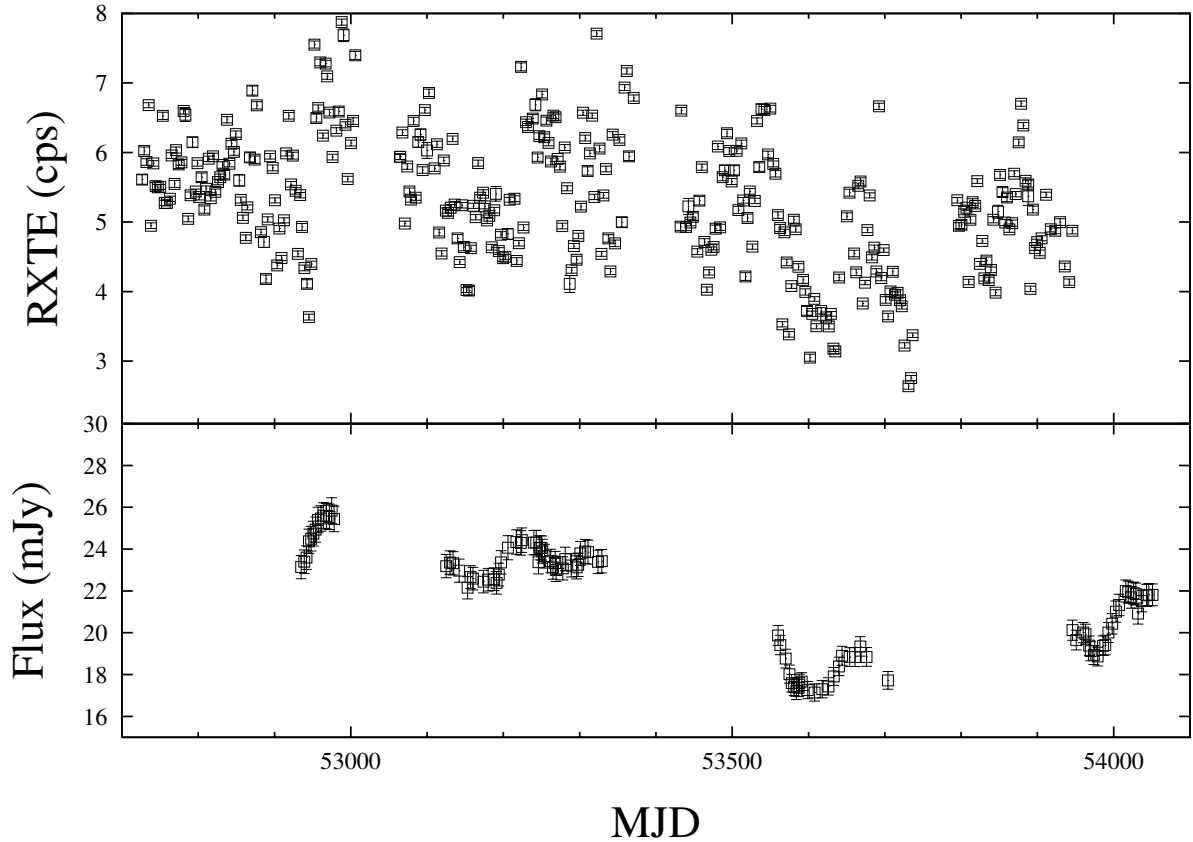}
\caption{X-ray (top) and R-band optical (bottom) light curves for Mkn~509.
Units are counts/sec/PCU and milli-Janskys for X-ray and optical data,
respectively.}
\end{figure}

\begin{figure}
\figurenum{2}
\plotone{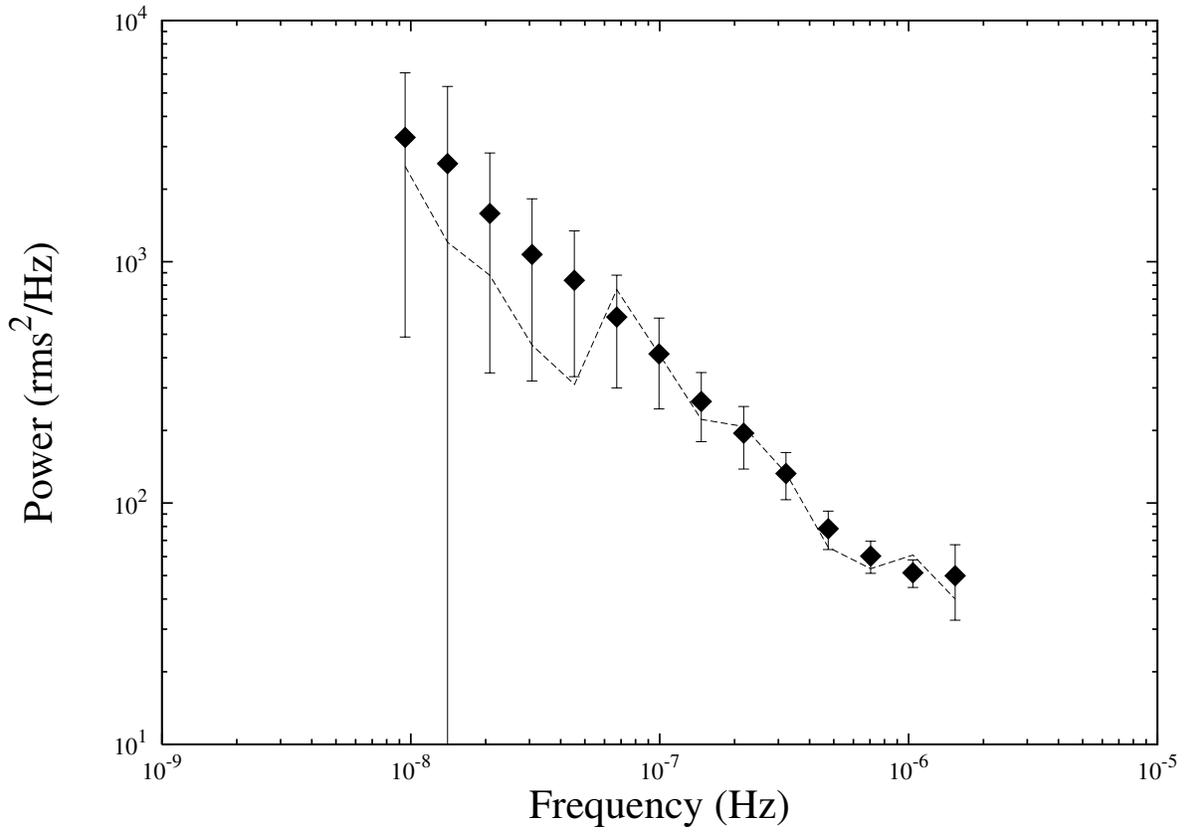}
\caption{X-ray power density spectrum.  Dashed line represents observed PDS,
while points with error bars represent simulation results.  Best fit model has
a high-frequency slope of $-2.0$, with a break frequency of $\log \nu_{\rm b} = -6.47$.}
\vspace{0.25in}
\end{figure}

\begin{figure}
\figurenum{3}
\includegraphics[angle=90,width=0.9\textwidth]{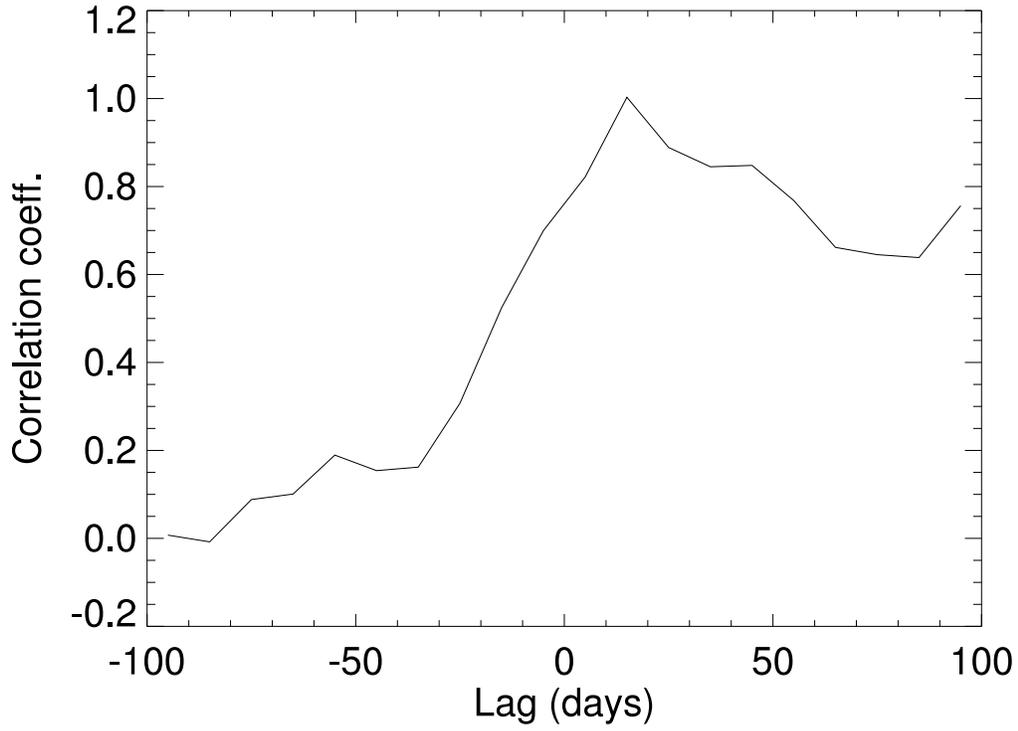}
\caption{Discrete correlation function for X-ray and optical light curves.
Positive lag indicates optical leading X-ray variations.  See text for a
discussion of errors.}
\end{figure}

\end{document}